\documentclass[conference, a4paper]{IEEEtran}
\IEEEoverridecommandlockouts

\makeatletter
\newcommand{\AddInputPath}[1]{%
  \ifx\input@path\@undefined
    \def\input@path{#1}
  \else
    \g@addto@macro{\input@path}{#1}
  \fi
}
\makeatother

\usepackage{shellesc}

\usepackage[font=footnotesize,caption=false]{subfig} 

\usepackage{relsize}

\usepackage[dvipsnames,svgnames]{xcolor}
\usepackage{array}
\usepackage{booktabs,tabularx}
\usepackage{multirow}
\usepackage[binary-units=true,per-mode=symbol,detect-mode=true]{siunitx}
\usepackage{graphicx}
\DeclareSIUnit \dBi {dBi}

\graphicspath{{images/}} 

\usepackage[T1]{fontenc}
\usepackage{textcomp}
\usepackage[utf8]{inputenc}
\usepackage[final]{microtype}
\usepackage{icomma}
\usepackage{xspace}

\usepackage[tbtags]{amsmath}
\usepackage{amssymb,amsfonts,bm}
\usepackage{mathtools} 
\usepackage{dsfont}
\usepackage{mathrsfs}
\usepackage{accents}
\usepackage{empheq}
\usepackage{nccmath}
\usepackage{balance}
\usepackage{setspace}

\usepackage{color}
\usepackage{calc}
\usepackage{tikz}
\usepackage{pgfplots,pgfplotstable}

\usepackage{pdftexcmds}
\makeatletter
\newcommand{\strequal}[2]{\pdf@strcmp{#1}{#2}==0}
\makeatother

\usepackage[capitalize]{cleveref}
\usepackage{refcount}

\usepackage[inline]{enumitem}
\usepackage{algorithm}
\usepackage{algpseudocode}
\makeatletter
\newcommand{\algmargin}{\the\ALG@thistlm}
\algblockx{StartIA}{EndIA}[1]{\textbf{Input} #1}[1]{\textbf{Return} #1}
\makeatother
\newlength{\whilewidth}
\algdef{SE}[parWHILE]{parWhile}{EndparWhile}[1]
  {\parbox[t]{\dimexpr\linewidth-\algmargin}{%
     \hangindent\whilewidth\strut\algorithmicwhile\ #1\ \algorithmicdo\strut}}{\algorithmicend\ \algorithmicwhile}%
\algnewcommand{\parState}[1]{\State%
  \parbox[t]{\dimexpr\linewidth-\algmargin}{\strut #1\strut}}

\makeatletter
\newcommand\fs@spaceruled{\def\@fs@cfont{\bfseries}\let\@fs@capt\floatc@ruled
  \def\@fs@pre{\vspace{.1in}\hrule height.9pt depth0pt \kern3pt}%
  \def\@fs@post{\kern3pt\hrule\relax}%
  \def\@fs@mid{\kern3pt\hrule\kern3pt}%
  \let\@fs@iftopcapt\iftrue}
\makeatother

\floatstyle{spaceruled}
\restylefloat{algorithm}

\usepackage{glossaries-prefix}
\usepackage{ifthen}
\usepackage[noadjust]{cite}
\usepackage{multibib}
\usepackage{dsfont}
\usepackage{comment}
\usepackage{todonotes}
\let\legacytodo\todo
\newcommand{\ruggedtodo}[2][]{\tikzexternaldisable\legacytodo[#1]{#2}\tikzexternalenable}
\renewcommand{\todo}[1]{\ruggedtodo[inline]{#1}}

\pgfplotsset{
    legend image with text/.style={
        legend image code/.code={%
            \node[anchor=center] at (0.3cm,0cm) {#1};
        }
    },
}

\makeatletter
\def\parsenode[#1]#2\pgf@nil{%
    \tikzset{label node/.style={#1}}
    \def\nodetext{#2}
}

\makeglossaries
\newacronym{cl_sia}{CL-SIA}{constant-length sparse incremental aggregation}
\newacronym{sia}{SIA}{sparse incremental aggregation}
\newacronym{soa}{SoA}{state-of-the-art}
\newacronym{mh}{MH}{multi-hop}
\newacronym{tcs}{TCS}{time correlated sparsification}
\newacronym{ia}{IA}{incremental aggregation}
\newacronym{sfl}{SFL}{satellite federated learning}
\newacronym{esa}{ESA}{European Space Agency}
\newacronym{tle}{TLE}{two-line element set}
\newacronym{ai}{AI}{artificial intelligence}
\newacronym{ann}{ANN}{artificial neural network}
\newacronym{jscc}{JSCC}{joint source-channel coding}
\newacronym{raan}{RAAN}{right ascension of the ascending node}
\newacronym{uav}{UAV}{unmanned aerial vehicle}
\newacronym{haps}{HAPS}{high-altitude platform station}
\newacronym{6g}{6G}{sixth generation}
\newacronym{cgr}{CGR}{contact graph routing}
\newacronym{dtn}{DTN}{delay-tolerant networking}
\newacronym{fl}{FL}{federated learning}
\newacronym{dl}{DL}{deep learning}
\newacronym{fedavg}{FedAvg}{federated averaging}
\newacronym{dml}{DML}{distributed ML}
\newacronym{ps}{PS}{parameter server}
\newacronym[prefix={an\space},prefixfirst={a~}]{ml}{ML}{machine learning}
\newacronym{sgd}{SGD}{stochastic gradient descent}
\newacronym{dsgd}{DSGD}{distributed stochastic gradient descent}
\newacronym{isl}{ISL}{inter-satellite link}
\newacronym{gsl}{GSL}{ground to satellite link}
\newacronym{gs}{GS}{ground station}
\newacronym{ecef}{ECEF}{earth-centered, earth-fixed}
\newacronym{eci}{ECI}{Earth-centered inertial}
\newacronym{ofdm}{OFDM}{orthogonal frequency-division multiplexing}
\newacronym{cp}{CP}{cyclic prefix}
\newacronym{los}{LoS}{line-of-sight}
\newacronym{leo}{LEO}{low earth orbit}
\newacronym{meo}{MEO}{medium earth orbit}
\newacronym{gso}{GSO}{geosynchronous orbit}
\newacronym{geo}{GEO}{geostationary}
\newacronym{eo}{EO}{Earth observation}
\newacronym{iot}{IoT}{Internet of Things}
\newacronym{irs}{IRS}{intelligent reflecting surface}
\newacronym{socp}{SOCP}{second-order cone program}
\newacronym{soc}{SOC}{second-order cone}
\newacronym{dsl}{DSL}{digital subscriber line}
\newacronym{wsee}{WSEE}{weighted sum energy efficiency}
\newacronym{mmwave}{mmWave}{millimeter wave}
\newacronym{dfg}{DFG}{Deutsche Forschungsgemeinschaft}
\newacronym{haec}{HAEC}{Highly Adaptive Energy-Efficient Computing}
\newacronym{hpc}{HPC}{High Performance Computing}
\newacronym{mac}{MAC}{multiple-access channel}
\newacronym{bc}{BC}{broadcast channel}
\newacronym{siso}{SISO}{single-input single-output}
\newacronym{simo}{SIMO}{single-input multiple-output}
\newacronym{miso}{MISO}{multiple-input single-output}
\newacronym{mimo}{MIMO}{multiple-input multiple-output}
\newacronym{af}{AF}{amplify-and-forward}
\newacronym{df}{DF}{decode-and-forward}
\newacronym{cf}{CF}{compress-and-forward}
\newacronym{mwrc}{MWRC}{multi-way relay channel}
\newacronym{dmmwrc}{DM-MWRC}{discrete memoryless multi-way relay channel}
\newacronym{pde}{PDE}{partial data exchange}
\newacronym{fde}{FDE}{full data exchange}
\newacronym{iid}{i.i.d.\@}{independent and identically distributed}
\newacronym{di}{DI} {difference of increasing}
\newacronym{dc}{DC}{difference of convex}
\newacronym{mm}{MM}{mixed monotonic}
\newacronym{mmp}{MMP}{mixed monotonic programming}
\newacronym{awgn}{AWGN}{additive white Gaussian noise}
\newacronym{wgn}{WGN}{white Gaussian noise}
\newacronym{awg}{AWG}{additive white Gaussian}
\newacronym{sic}{SIC}{successive interference cancellation}
\newacronym{snr}{SNR}{signal-to-noise ratio}
\newacronym{sinr}{SINR}{signal to interference plus noise ratio}
\newacronym{inr}{INR}{interference to noise ratio}
\newacronym{zf}{ZF}{zero-forcing}
\newacronym{mrt}{MRT}{maximum ratio transmission}
\newacronym{mmse}{MMSE}{minimum mean square error}
\newacronym{sud}{SUD}{single user decoding}
\newacronym{dof}{DoF}{degrees of freedom}
\newacronym{gdof}{GDoF}{generalized degrees of freedom}
\newacronym{nnc}{NNC}{noisy network coding}
\newacronym{dmn}{DMN}{discrete memoryless network}
\newacronym{csi}{CSI}{channel state information}
\newacronym{pmf}{pmf}{probability mass function}
\newacronym{dmic}{DM-IC}{discrete memoryless interference channel}
\newacronym{ic}{IC}{interference channel}
\newacronym{gic}{GIC}{Gaussian interference channel}
\newacronym{if}{IF}{interference}
\newacronym{ee}{EE}{energy efficiency}
\newacronym{gee}{GEE}{global energy efficiency}
\newacronym{tin}{TIN}{treating interference as noise}
\newacronym{snd}{SND}{simultaneous non-unique decoding}
\newacronym{sd}{SD}{simultaneous decoding}
\newacronym{hk}{HK}{Han-Kobayashi}
\newacronym{rs}{RS}{rate splitting}
\newacronym{rf}{RF}{radio frequency}
\newacronym{pa}{PA}{power amplifier}
\newacronym{lna}{LNA}{low noise amplifier}
\newacronym{lo}{LO}{local oscillator}
\newacronym{adc}{ADC}{analog-to-digital converter}
\newacronym{dac}{DAC}{digital-to-analog converter}
\newacronym{dsp}{DSP}{digital signal processing}
\newacronym{brd}{BRD}{best response dynamics}
\newacronym{br}{BR}{best response}
\newacronym{ne}{NE}{Nash equilibrium}
\newacronym{lhs}{LHS}{left-hand side}
\newacronym{rhs}{RHS}{right-hand side}
\newacronym{ran}{RAN}{radio access network}
\newacronym{qos}{QoS}{Quality of Service}
\newacronym{ngmn}{NGMN}{Next Generation Mobile Networks}
\newacronym{cap}{CAP}{Capacity Adaptation}
\newacronym{bwa}{BW}{Bandwidth Adaptation}
\newacronym{prb}{PRB}{physical resource block}
\newacronym{se}{SE}{spectral efficiency}
\newacronym{tp}{TP}{throughput}
\newacronym{bs}{BS}{base station}
\newacronym{ue}{UE}{user equipment}
\newacronym{mop}{MOP}{multi-objective optimization problem}
\newacronym{gda}{GDA}{generalized Dinkelbach's algorithm}
\newacronym{midcp}{MIDCP}{mixed integer disciplined convex programming}
\newacronym{lp}{LP}{linear program}
\newacronym{brb}{BRB}{branch reduce and bound}
\newacronym{bb}{BB}{branch and bound}
\newacronym{sit}{SIT}{successive incumbent transcending}
\newacronym{oma}{OMA}{orthogonal multiple access}
\newacronym{noma}{NOMA}{non-orthogonal multiple access}
\newacronym{wlog}{w.l.o.g.\@}{without loss of generality}
\newacronym{lsc}{l.s.c.\@}{lower semi-continuous}
\newacronym{usc}{u.s.c.\@}{upper semi-continuous}
\newacronym{kkt}{KKT}{Karush-Kuhn-Tucker}
\newacronym{ptp}{PTP}{point-to-point}

\glsenableentrycount

\usetikzlibrary{positioning}
\usetikzlibrary{calc}
\usetikzlibrary{math}
\usetikzlibrary{fit}
\usetikzlibrary{intersections}
\usetikzlibrary{decorations.pathreplacing}
\usetikzlibrary{decorations.markings}
\usetikzlibrary{3d,angles}
\usetikzlibrary{arrows.meta}

\pgfdeclarelayer{background}
\pgfsetlayers{background,main}

\usetikzlibrary{external}

\usepgfplotslibrary{colorbrewer}

\tikzset{
	small1/.style={fill=DeepPink},
	small2/.style={fill=DeepSkyBlue},
	small3/.style={fill=MediumSpringGreen},
	ps/.style={fill=Gold},
	link/.style = {semithick},
	plane/.style={plane origin={(#1,0,0)}, plane x = {(#1,0,1)}, plane y = {(#1,1,0)}, rotate around y = -9, canvas is plane}
}

\tikzset{
	antenna/.pic={
		\draw[thick] (0,0) -- ++(120:2mm) -- ++(0:2mm) -- cycle -- (0,-1.5mm);
	}
}


\crefname{equation}{}{}
\crefrangeformat{equation}{(#3#1#4)--(#5#2#6)}
\crefmultiformat{equation}{(#2#1#3)}{ and~(#2#1#3)}{, (#2#1#3)}{, (#2#1#3)}
\crefrangemultiformat{equation}{#3(#1)#4--#5(#2)#6}{, #3(#1)#4--#5(#2)#6}{, #3(#1)#4--#5(#2)#6}{, #3(#1)#4--#5(#2)#6}
\crefrangeformat{algorithm}{Algorithms~#3#1#4--#5#2#6}

\undef\mod
\DeclareMathOperator\mod{mod}

\let\vec\bm

\allowdisplaybreaks[3]

\DeclareSIUnit \dBm {dBm}
\DeclareSIUnit \dBW {dBW}
\DeclareSIUnit \bpcu {bpcu}

\DeclareFontFamily{U}{mathx}{\hyphenchar\font45}
\DeclareFontShape{U}{mathx}{m}{n}{
      <5> <6> <7> <8> <9> <10>
      <10.95> <12> <14.4> <17.28> <20.74> <24.88>
      mathx10
      }{}
\DeclareSymbolFont{mathx}{U}{mathx}{m}{n}
\DeclareMathSymbol{\bigtimes}{1}{mathx}{"91}

\pgfplotscreateplotcyclelist{default}{%
	blue,mark=*\\%
	red,mark=star\\%
	teal,mark=square*\\%
	brown!60!black,mark=otimes*\\%
}

\definecolor{plot1}{RGB}{228,26,28}
\definecolor{plot2}{RGB}{55,126,184}
\definecolor{plot3}{RGB}{77,175,74}
\definecolor{plot4}{RGB}{152,78,163}
\definecolor{plot5}{RGB}{255,127,0}
\definecolor{plot6}{RGB}{166,86,40}
\tikzstyle{fedsatschedule}=[plot1]
\tikzstyle{fedsat}=[plot2]
\tikzstyle{fedisl}=[plot3]
\tikzstyle{fedavg}=[plot4]
\tikzstyle{fedasync1}=[plot5]
\tikzstyle{fedasync2}=[plot6]

\newcolumntype{P}[1]{>{\centering\arraybackslash}p{#1}}

\ifCLASSOPTIONdraftcls
\AtBeginEnvironment{figure}{}
\fi

\AtBeginEnvironment{algorithmic}{\footnotesize}

\begin{document}
\bstctlcite{IEEEexample:BSTcontrol}
\title{Sparse Incremental Aggregation in Satellite Federated Learning}

\author{\IEEEauthorblockN{ Nasrin Razmi\IEEEauthorrefmark{1}\IEEEauthorrefmark{2}, Sourav Mukherjee\IEEEauthorrefmark{1}\IEEEauthorrefmark{2},  Bho Matthiesen\IEEEauthorrefmark{1}\IEEEauthorrefmark{2}, Armin Dekorsy\IEEEauthorrefmark{1}\IEEEauthorrefmark{2}, Petar Popovski\IEEEauthorrefmark{3}\IEEEauthorrefmark{1}}
	\IEEEauthorblockA{\IEEEauthorrefmark{1}University of Bremen, Department of Communications Engineering, Germany\\\IEEEauthorblockA{\IEEEauthorrefmark{2}Gauss-Olbers Space Technology
Transfer Center, University of Bremen, Germany\\\IEEEauthorrefmark{3}Aalborg University, Department of Electronic Systems, Denmark\\ email: \{razmi, mukherjee, dekorsy, matthiesen\}@ant.uni-bremen.de, petarp@es.aau.dk}}
\thanks{
This work is supported by the German Research Foundation (DFG) under Grant EXC 2077 (University Allowance).
}%
}
\maketitle
\begin{abstract}
This paper studies Federated Learning (FL) in low Earth orbit (LEO) satellite constellations, where satellites are connected via intra-orbit inter-satellite links (ISLs) to their neighboring satellites. During the FL training process, satellites in each orbit forward gradients from nearby satellites, which are eventually transferred to the parameter server (PS). To enhance the efficiency of the FL training process, satellites apply in-network aggregation, referred to as incremental aggregation. In this work, the gradient sparsification methods from \cite{Sourav2024sparse} are applied to satellite scenarios to improve bandwidth efficiency during incremental aggregation. The numerical results highlight an increase of over $4 \times$ in bandwidth efficiency as the number of satellites in the orbital plane increases.
\end{abstract}
\glsresetall

\section{Introduction}
Megaconstellations of low Earth orbit (LEO) satellites have become an integral part to a variety of applications including global broadband access, Earth monitoring, and space exploration missions. The massive amount of data produced by these satellites—particularly high-resolution hyperspectral images—creates significant challenges in transmitting them back to the Earth. These challenges are exacerbated by the limitations of available bandwidth and the need to meet stringent latency requirements \cite{giuffrida2020cloudscout, Leyva2020LEO}.

\cGls{sfl} has recently developed as an promising technology to address the above challenges \cite{Matthiesen2022FL, Razmi2024On-Board, Chen2022Sat}. Unlike traditional approaches, \cgls{sfl} leverages distributed machine learning by enabling satellites to collaboratively train \cgls{ml} models without exchanging raw data. In this approach, each satellite trains the model locally using its own dataset and the global model parameters received from the central \cgls{ps}, and then transmits the updated model parameters to the \cgls{ps} for aggregation. However, due to satellite orbital motion, the \cgls{ps} must wait a long time to receive model updates, making conventional \cgls{fl} impractical for satellite constellations. 

To alleviate the impractical delay in the model convergence, a practical asynchronous \cgls{fl} approach was presented in \cite{Razmi2022Ground}, where each satellite functions as a separate node, communicating with a \cgls{ps} located at a \cgls{gs} on Earth for model aggregation. Another asynchronous \cgls{fl} is presented in \cite{wu2023fedgsm} to mitigate model staleness, leading to faster model convergence. Model accuracy of \cite{Razmi2022Ground}  is further improved in \cite{razmi2022scheduling} by leveraging the predictability of satellite connections to the \cgls{ps} to schedule the transmissions of model parameters. 

Further, to improve the model accuracy of \cgls{sfl} and address the long delays, \cite{Razmi2024On-Board} proposed to aggregate the updated model parameters inside each orbital plane and then transmitting the aggregated results to the \cgls{ps}. To this end, the satellites forward the updated parameters using their intra-orbit \cglspl{isl}. This approach allows the necessary information from all satellites within an orbit to be accessed through a single satellite that remains visible to the \gls{gs}. Moreover, inter-orbit \cglspl{isl} are utilized in \cite{Shi2024SatEdge, ZHou2024Decentralized, Zhai2024FedLEO} to accelerate convergence.

In \cite{Razmi2024On-Board}, to improve the bandwidth efficiency for model aggregation, each satellite applies gradient sparsification methods \cite{Aji2017, alistarh2018convergence}. In the approach proposed in \cite{Razmi2024On-Board}, each satellite transmits the most effective elements of its parameters and the corresponding indices, along with the received elements and their indices to its neighboring satellite. For smaller sparsification ratios, the index supports for effective elements of the satellites are almost uncorrelated; therefore, making the aggregation inefficient and increasing communication budget with each hop~\cite{Razmi2024On-Board}. The growing communication budget may become a bottleneck in transmission through the bandwidth-limited \cglspl{isl}. 

In this paper, building upon our prior research \cite{Razmi2024On-Board}, we apply the sparsification approaches proposed in \cite{Sourav2024sparse} to the satellite constellations connected with intra-orbit \cglspl{isl} to improve the bandwidth efficiency while maintaining efficient communication. The structure of the paper is as follows: first, we introduce the system model in \cref{sec: systemmod}. Next, \cgls{fl} with intra-orbit \cglspl{isl} and sparsified gradient transmissions are discussed in \cref{sec:fl_intra} and \cref{sec:sparsIA}, respectively. Finally, numerical results and conclusions are presented in \cref{sec:numeval} and \cref{sec:Conclusions}.


\section{System Model} \label{sec: systemmod}

\subsection{Satellite Constellations} \label{sec: satcons}
We consider a satellite constellation with a total of $P$ orbital planes, where each plane $p$ contains $K_p$ satellites. The set of all satellites in the constellation is denoted as $\mathcal K = \bigcup_{p = 1}^P \mathcal K_p = \{ k_{1,1},\dots,k_{P,K_P} \}$, where the total number of satellites is $K = \sum_{p=1}^P K_p$. Satellites in orbit $p$ move at a speed determined by $v_p = \sqrt{\frac{\mu}{h_p+r_E}}\,\si{\meter/\second}$, where $\mu = 3.98 \times 10^{14}\,\si{\meter^3/\second^2}$ is the geocentric gravitational constant, $h_p$ is the orbit altitude, and $r_E = 6371\,\si{\km}$ is the Earth's radius. The orbital period of the satellites is calculated as $T_p = \frac{2\pi(r_E + h_p)}{v_p}$.

\subsection{Computation Model} \label{sec: comp}

Satellites in the constellation participate in training an \cgls{fl} algorithm  using the FedAvg method \cite{mcmahan2017communication}. To this end, each satellite $k$ utilizes its collected data set $\mathcal{D}_k$ and trains a \cgls{ml} model. In the training process, satellites solve the optimization problem 

\begin{equation} \label{eq:global_FL_problem}
	F(\vec{w}) = \min\nolimits_{\vec w}\enskip {\sum_{k=1}^K} \frac{D_k}{D} F_k(\vec {w}),
\end{equation}
where $D_k = |\mathcal{D}_k|$, $D = \sum_{k=1}^{K} D_k$ and the local loss function $F_k(\vec {w})$ is defined as

\begin{equation} \label{eq:local_FL_problem}
   F_k(\vec {w})= \frac{1}{D_k} \sum_{\vec{x} \in \mathcal{D}_k} f(\vec{w},\vec{x}), 
\end{equation}
with $f(\vec{w},\vec{x})$ as the per-sample loss function. The training process is coordinated by the \cgls{ps} over $N$ iterations to solve \cref{eq:global_FL_problem}. We consider a \cgls{gs} functioning as the \cgls{ps}.

	\begin{figure*}[t]
 \label{fig: incremental_aggregation}
		\centering
		\begin{tikzpicture}[every node/.style = {font=\scriptsize} ]
			\begin{scope}[scale = .3]
				\node[anchor=south, font = \footnotesize] at (6,1) {Satellite 1};
				\draw[fill=yellow] (0,0) ++(0,1) foreach \x in {0,1,2,...,11} {++(0,-1) rectangle ++(1,1)};

				\node[anchor=south, font = \footnotesize] at (23,1) {Satellite 2};
				\draw[fill=red] (17,0) ++(0,1) foreach \x in {0,1,2,...,11} {++(0,-1) rectangle ++(1,1)};

\draw[thick] (16,-2.5) -- ++(13.5,0);
				\node [anchor = center] at (16.5,-1.5) {+};				

				\draw[fill=yellow] (17,-2) ++(0,1) foreach \x in {0,1,2,...,11} {++(0,-1) rectangle ++(1,1)};

\fill[yellow] foreach \x in {0,1,...,10,11} {(17+\x,-4) -| ++(1,1) --cycle};

\fill[red] foreach \x in {0,1,...,10,11} {(17+\x,-4) |- ++(1,1) --cycle};

\draw (17,-4) ++(0,1) foreach \x in {0,1,2,...,11} {++(0,-1) rectangle ++(1,1)};

				\node[anchor=south, font = \footnotesize] at (40,1) {Satellite 3};
				\draw[fill=green] (34,0) ++(0,1) foreach \x in {0,1,2,...,11} {++(0,-1) rectangle ++(1,1)};

\fill[yellow] foreach \x in {0,1,...,10,11} {(34+\x,-2) -| ++(1,1) --cycle};

\fill[red] foreach \x in {0,1,...,10,11} {(34+\x,-2) |- ++(1,1) --cycle};

\draw[thick] (33.5,-2.5) --(46.7,-2.5);
				\node [anchor = center] at (33.5,-1.5) {+};

\draw (34,-2) ++(0,1) foreach \x in {0,1,2,...,11} {++(0,-1) rectangle ++(1,1)};

\draw (34,-4) ++(0,1) foreach \x in {0,1,2,...,11} {++(0,-1) rectangle ++(1,1)};

\fill[yellow] foreach \x in {0,1,...,10,11} {(34+\x,-4) rectangle ++(1,1)};

\fill[red] foreach \x in {0,1,...,10,11} {(34+\x,-4) |- ++(1,1) --cycle};

\fill[green] foreach \x in {0,1,...,10,11} {(34+\x,-3) -| ++(1,-1) --cycle};


				\draw[semithick,-latex] (12.25,0.5) -- ++(1.875,0) -- ++(0,-2) -- ++(1.875,0);
				\draw[thick,-latex] (29.45,-3.5) -- ++(1.875,0) -- ++(0,2.3) -- ++(1.875,0);
			\end{scope}
		\end{tikzpicture}
  \caption{The incremental aggregation method is applied across three adjacent satellites, with Satellite 1 being the farthest from the sink.}
	\end{figure*}
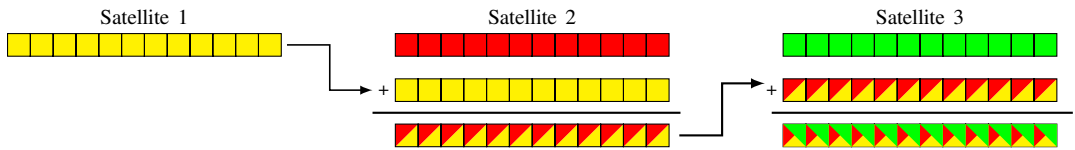

For the $n$th iteration, each satellite $k$ after receiving the global model parameters $\vec{w}^n$,  performs $I$ local steps of mini-batch stochastic gradient descent to obtain $\vec{w}_k^{n,I}$, as described in \cref{alg:local_learn_steps} \cite{Razmi2022Ground}. 
\begin{algorithm}[t!] 
	\caption{Satellite learning procedure} \label{alg:Satellite SGD Procedure}
	\begin{algorithmic}[1]
		\Procedure{SatLearnProc}{$\vec{w}^{n}$}
			\State \textbf{initialize} $\vec{w}_k^{n,0} = \vec{w}^{n}, \quad i = 0$, \quad learning rate $\eta$ 
        \label{alg:ssgd:init}
			\For {$I$ epochs} \label{alg:ssgd:batchstart}
				\Comment $I$ epochs of mini-batch \cgls{sgd}
				\State $\tilde{\mathcal D}_k \gets $ Randomly shuffle $\mathcal D_k$
				\State $\mathscr B \gets $ Partition $\tilde{\mathcal D}_k$ into mini-batches of size $B$
				\For {each batch $\mathcal B\in\mathscr B$}
				\State $\vec{w}_k^{n,i} \gets \vec{w}_k^{n,i} - \frac{\eta}{|\mathcal B|} \nabla_{\vec{w}} \left(\sum_{\vec x\in\mathcal B} f(\vec x, \vec w) \right)$				
				\EndFor
                        \State $i \gets i + 1$
			\EndFor \label{alg:ssgd:batchend}
		      \label{alg:ssgd:compress}
			\State \Return {$\vec{w}^{n,I}_{k}$} \label{alg:local_learn_steps}
		\EndProcedure
	\end{algorithmic}
\end{algorithm}
Then, the satellite transmits the gradients $\vec{g}_{k}^{n}$ to the \cgls{gs}, where $\vec{g}_{k}^{n}$ is defined as
\begin{equation} \label{eq: grad}
\vec{g}_{k}^{n} = \vec{w}_{k}^{n,I} - \vec{w}^{n}.
\end{equation}
The \cgls{gs} then aggregates $\vec{g}_{k}^{n}$, and updates the global model parameters as 
\begin{equation} \label{eq: global update}
    \vec{w}^{n+1} = \vec{w}^{n} + \sum_{k=1}^{K} \frac{D_k}{D} \vec{g}_{k}^{n},
\end{equation}
Afterwards, the \cgls{gs} transmits $\vec{w}^{n+1}$ back to the satellites for the next iteration.

\begin{figure*}[t]
          \tikzsetnextfilename{aggregation_sparsification_explain_with_three_satellite_ag}
		\centering
		\begin{tikzpicture}[every node/.style = {font=\scriptsize} ]
			\begin{scope}[scale = .3]
				\node[anchor=south, font = \footnotesize] at (6,1) {Satellite 1};

    \draw[fill=yellow] (0,0) ++(0,1) foreach \x in {0,1,2,...,11} {++(0,-1) rectangle ++(1,1)};
				\draw[-latex] (6,-.25) --node[right] {$\mathsf{Top}_3$} ++(0,-1.5);

    \fill[yellow] foreach \x in {3,7,9} {(0+\x,-3) rectangle ++(1,1)};

    \draw (0,-3) ++(0,1) foreach \x in {0,1,2,...,11} {++(0,-1) rectangle ++(1,1)};

				\node[anchor=south, font = \footnotesize] at (23,1) {Satellite 2};
				\draw[fill=red] (17,0) ++(0,1) foreach \x in {0,1,2,...,11} {++(0,-1) rectangle ++(1,1)};
				\draw[-latex] (23,-.25) --node[right] {$\mathsf{Top}_3$} ++(0,-1.5);
				\fill[red] foreach \x in {1,3,11} {(17+\x,-3) rectangle ++(1,1)};
				\draw (17,-3) ++(0,1) foreach \x in {0,1,2,...,11} {++(0,-1) rectangle ++(1,1)};

				\fill[yellow] foreach \x in {3,7,9} {(17+\x,-4.5) rectangle ++(1,1)};
				\draw (17,-4.5) ++(0,1) foreach \x in {0,1,2,...,11} {++(0,-1) rectangle ++(1,1)};

				\draw[thick] (16,-4.9) -- ++(13.5,0);
				\node [anchor = center] at (16.5,-4) {+};

				\fill[yellow] foreach \x in {3,7,9} {(17+\x,-6.3) rectangle ++(1,1)};
				\fill[red] foreach \x in {1,11} {(17+\x,-6.3) rectangle ++(1,1)};
				\fill[red] (17+3,-6.3) -| ++(1,1) --cycle;
				\draw (17,-6.3) ++(0,1) foreach \x in {0,1,2,...,11} {++(0,-1) rectangle ++(1,1)};

				\node[anchor=south, font = \footnotesize] at (40,1) {Satellite 3};
				\draw[fill=green] (34,0) ++(0,1) foreach \x in {0,1,2,...,11} {++(0,-1) rectangle ++(1,1)};
				\draw[-latex] (40,-.25) --node[right] {$\mathsf{Top}_3$} ++(0,-1.5);
				\fill[green] foreach \x in {1,3,10} {(34+\x,-3) rectangle ++(1,1)};
				\draw (34,-3) ++(0,1) foreach \x in {0,1,2,...,11} {++(0,-1) rectangle ++(1,1)};

				\fill[yellow] foreach \x in {3,7,9} {(34+\x,-4.5) rectangle ++(1,1)};
				\fill[red] foreach \x in {1,11} {(34+\x,-4.5) rectangle ++(1,1)};
				\fill[red] (34+3,-4.5) -| ++(1,1) --cycle;
				\draw (34,-4.5) ++(0,1) foreach \x in {0,1,2,...,11} {++(0,-1) rectangle ++(1,1)};

				\draw[thick] (33,-4.9) -- ++(13.5,0);
				\node [anchor = center] at (33.5,-4) {+};

				\fill[yellow] foreach \x in {3,7,9} {(34+\x,-6.3) rectangle ++(1,1)};
				\fill[red] foreach \x in {1,11} {(34+\x,-6.3) rectangle ++(1,1)};
				\fill[green] foreach \x in {10} {(34+\x,-6.3) rectangle ++(1,1)};
				\fill[red] (34+3,-6.3) -| ++(1,1) --cycle;
				\fill[green] (34+3,-5.3) -| ++(1,-1) --cycle;
				\fill[green] (34+1,-5.3) -| ++(1,-1) --cycle;
				\draw (34,-6.3) ++(0,1) foreach \x in {0,1,2,...,11} {++(0,-1) rectangle ++(1,1)};

				\draw[semithick,-latex] (12.25,-2.5) -- ++(1.875,0) -- ++(0,-1.5) -- ++(1.875,0);
				\draw[thick,-latex] (29.25,-5.8) -- ++(1.875,0) -- ++(0,1.8) -- ++(1.875,0);
			\end{scope}
		\end{tikzpicture}
  \caption{ The Sparse incremental aggregation method is applied across three adjacent satellites, with Satellite 1 being the farthest from the sink.}
  \label{fig: Sparse_IA}
	\end{figure*}
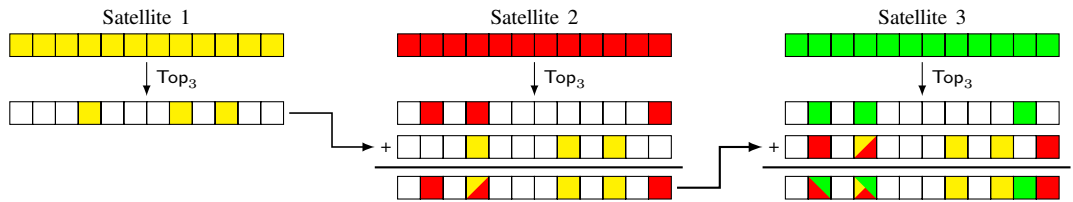

\subsection{Communication Model}\label{sec: comm}
To enable gradient transmission, each satellite in the constellation is equipped with three communication devices: one for the communication with the \cgls{gs} and the other two for the intra-orbit \cglspl{isl}. In each plane, each satellite connects with two of its nearest orbital neighbors, establishing a ring network. Further, the communication with the \cgls{gs} is only possible when the Earth does not obstruct the \cgls{los}.

We model the channel between two satellites $k$ and $i$ as a complex Gaussian channels, where the maximum achievable data rate is $r(k,i)=B \log_2(1+\mathrm {SNR}(k,i))$, with $B$ representing the allocated bandwidth, and the \cgls{snr}~(SNR) given as \cite{Leyva2021ISL, ippolito2017satellite}
\begin{equation} \label{eq: SNR}
    \mathrm {SNR}(k,i) = \frac{P_t G_k(i) G_i(k) }{N_0  L(k,i)}.
\end{equation}
Further, the $\mathrm {SNR}(k,i)$ is zero, if there is no \cgls{los}. Here, $G_k(i)$ denotes the average antenna gain of satellite $k$ towards satellite $i$, and $N_0= k_B T B$ is the total noise power, where the Boltzmann constant is $k_B = 1.380649 {\times }10^{-23}\,\si{\joule/\kelvin}$ and $T$ is the receiver noise temperature. Additionally, $P_t$ is the transmitted power at the satellite, and
the free space path loss $L(k,i)$ is defined as
\begin{equation} \label{eq: loss}
L(k,i)=\left(\frac{4\pi f_c d(k,i)}{c}\right)^2,
\end{equation}
where $f_c$ is the carrier frequency, $c$ is the speed of light, and $d(k,i)$ is the distance between satellite $k$ and $i$. We assume fixed-rate transmission links with the minimum rate possible in the links. 

\section{Federated Learning using Intra-orbit ISLs} \label{sec:fl_intra}

For \cgls{fl} without the intra-orbit \cglspl{isl}, each satellite must connect two times to the \cgls{gs} during each iteration $n$: first, to receive the global parameters $\vec{w}^n$ and then, to transmit back the updated gradient $\vec{g}_k^{n}$. Both of these transmissions require \cgls{los} between \cgls{gs} and the satellite. However, due to the satellite’s motion, \cgls{los} to the \cgls{gs} lasts for a limited time during each orbital period $T_p$. Most of the time, the \cgls{los} is obstructed by the Earth.
Therefore, the \cgls{gs} must wait until each satellite is visible (\cgls{los}) to transmit the global parameters $\vec{w}^n$, and then wait again for the satellite to be visible to receive the updated gradient $\vec{g}_k^{n}$. This introduces significant delays in model training and can lead to outdated data in time-critical applications.

Long delays can be minimized by enabling neighboring satellites to communicate through intra-orbit \cglspl{isl} and by strategically scheduling transmissions to leverage the predictable paths of satellite orbits. This approach, termed \textit{incremental aggregation}, operates as follows: in each iteration \( n \), the satellite with \cgls{los} to the \cgls{gs} receives the global model parameters \( \vec{w}^n \). These parameters are then distributed to other satellites within the orbit via intra-orbit \cglspl{isl}. Another satellite with optimal visibility relays the updated aggregated gradients \( \vec{g}_p^n = \sum_{k=1}^{K_p} D_k \vec{g}_k^n \) back to the \cgls{gs}. Each iteration \( n \) for an orbit is therefore structured into three distinct phases: parameter distribution, computation, and aggregation.

In the parameter distribution phase, one satellite of the orbit $p$ which is selected for its optimal visibility to the \cgls{gs}, referred to as the source, receives $\vec{w}^n$ at time $t$. The source satellite then identifies another satellite in the same orbit as sink, which delivers the aggregated updates to the \cgls{gs}. The source selects sink satellite based on the computation time of each satellite, total aggregation time, and the expected visibility of the satellites. Once the source has identified the sink, it transmits $\vec{w}^n$, along with the sink’s ID, to its neighboring satellites in both directions. The neighboring satellites then relay the received $\vec{w}^n$ and the ID of the sink to the next satellite in sequence, continuing until all satellites have received the packet.

After forwarding the received packet, each satellite $k$ begins its training process as in  \cref{alg:local_learn_steps}. The aggregation phase starts after training phase. In this phase, the satellites farthest from the sink initiate the process by transmitting their updated gradients to nearest neighboring satellite, toward the sink. This process continues in both directions along the ring network \cite{Razmi2024On-Board, Razmi2022a}.

To illustrate the incremental aggregation method with more details, we assume satellite $k-1$ is the farthest satellite from the sink. This satellite first calculates the shortest path to the sink based on the sink ID received during distribution phase. Once the shortest path is determined, the satellite  transmits its gradients, scaled by its data size, as $\vec \gamma_{k-1}^n \leftarrow D_{k-1}\vec{g}_{k-1}^n$, to a neighboring satellite $k$, which is chosen based on the shortest path to the sink. Note that the shortest path selected by the satellite $k-1$ involving satellite $k$, is also the shortest path for the satellite $k$ towards the sink. Therefore, if the satellite $k$ performs shortest path search, it will also come up with the same path. The satellite $k$ then aggregates the received parameters with its own gradients as $\vec \gamma_{k}^n \leftarrow  D_{k}\vec{g}_{k}^n + \vec \gamma_{k-1}^n$, and transmits this to the next satellite, $k+1$. These steps continue until the sink receives the aggregated gradients from both directions. 

The key advantage of incremental aggregation is that it maintains a constant outgoing data size. Each satellite transmits data of size $n_d \omega$, where $n_d$ denotes the gradient dimension and $\omega$ represents the storage size for a single gradient entry. In each iteration, the total data transmitted within an orbital plane $p$ is $K_p n_d \omega$. However, when training models with large $n_d$ in satellite mega-constellations, the data transmission size can become a bottleneck, especially with bandwidth-limited \cglspl{isl}.

\section{Sparse Transmission} \label{sec:sparsIA}

To improve the bandwidth efficiency of incremental aggregation, an effective approach is to compress gradients into a sparser representation. Among various compression techniques, $\text{Top}_Q$ sparsification is a popular choice due to its strong performance. In the Top$_Q$ approach, the gradient 
 is converted into a sparse vector, retaining only the $Q$ largest-magnitude elements~\cite{alistarh2018convergence}. Thus, combining Top$_Q$ sparsification with incremental aggregation can significantly reduce the overall communication budget, making it particularly advantageous for satellite constellations.

	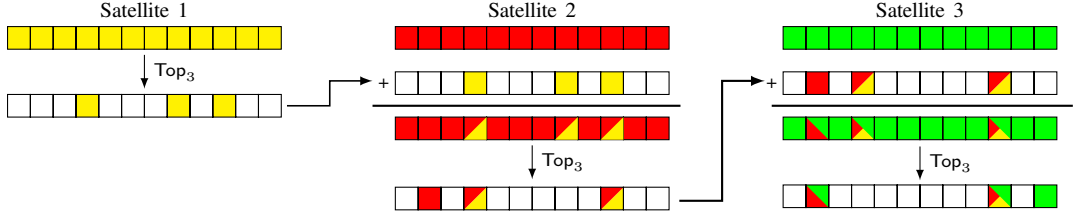
\begin{figure*}[t]
          \tikzsetnextfilename{aggregation_sparsification_explain_with_three_satellite_ag}
		\centering
		\begin{tikzpicture}[every node/.style = {font=\scriptsize} ]
			\begin{scope}[scale = .3]
				\node[anchor=south, font = \footnotesize] at (6,1) {Satellite 1};
				\draw[fill=yellow] (0,0) ++(0,1) foreach \x in {0,1,2,...,11} {++(0,-1) rectangle ++(1,1)};
				\draw[-latex] (6,-.25) --node[right] {$\mathsf{Top}_3$} ++(0,-1.5);
				\fill[yellow] foreach \x in {3,7,9} {(0+\x,-3) rectangle ++(1,1)};
				\draw (0,-3) ++(0,1) foreach \x in {0,1,2,...,11} {++(0,-1) rectangle ++(1,1)};

				\node[anchor=south, font = \footnotesize] at (23,1) {Satellite 2};
				\draw[fill=red] (17,0) ++(0,1) foreach \x in {0,1,2,...,11} {++(0,-1) rectangle ++(1,1)};

\draw[thick] (16,-2.5) -- ++(13.5,0);
				\node [anchor = center] at (16.5,-1.5) {+};				

				\fill[yellow] foreach \x in {3,7,9} {(17+\x,-2) rectangle ++(1,1)};
				\draw (17,-2) ++(0,1) foreach \x in {0,1,2,...,11} {++(0,-1) rectangle ++(1,1)};

\draw[fill=red] (17,-4) ++(0,1) foreach \x in {0,1,2,...,11} {++(0,-1) rectangle ++(1,1)};

\fill[yellow] foreach \x in {3,7, 9} {(17+\x,-4) -| ++(1,1) --cycle};

\draw[-latex] (23,-4.25) --node[right] {$\mathsf{Top}_3$} ++(0,-1.5);

				\fill[red] foreach \x in {1} {(17+\x,-7.1) rectangle ++(1,1)};

\fill[yellow] foreach \x in {3,9} {(17+\x,-7.1) -| ++(1,1) --cycle};

\fill[red] foreach \x in {3,9} {(17+\x,-7.1) |- ++(1,1) --cycle};

				\draw (17,-7.1) ++(0,1) foreach \x in {0,1,2,...,11} {++(0,-1) rectangle ++(1,1)};

				\node[anchor=south, font = \footnotesize] at (40,1) {Satellite 3};
				\draw[fill=green] (34,0) ++(0,1) foreach \x in {0,1,2,...,11} {++(0,-1) rectangle ++(1,1)};

\fill[red] foreach \x in {1} {(34+\x,-2) rectangle ++(1,1)};

\draw[thick] (33.5,-2.5) --(46.7,-2.5);
				\node [anchor = center] at (33.5,-1.5) {+};

\fill[yellow] foreach \x in {3,9} {(34+\x,-2) -| ++(1,1) --cycle};

\fill[red] foreach \x in {3,9} {(34+\x,-2) |- ++(1,1) --cycle};

				\draw (34,-2) ++(0,1) foreach \x in {0,1,2,...,11} {++(0,-1) rectangle ++(1,1)};

\fill[green] foreach \x in {0,2,4,5,6,7,8,10,11} {(34+\x,-4) rectangle ++(1,1) };

\fill[red] foreach \x in {1} {(34+\x,-4) rectangle ++(1,1)};

\fill[green] foreach \x in {1} {(34+\x,-3) -| ++(1,-1) --cycle};

\fill[yellow] foreach \x in {3,9} {(34+\x,-4) rectangle ++(1,1)};

\fill[red] foreach \x in {3,9} {(34+\x,-4) |- ++(1,1) --cycle};

\fill[green] foreach \x in {3,9} {(34+\x,-3) -| ++(1,-1) --cycle};

\draw (34,-4) ++(0,1) foreach \x in {0,1,2,...,11} {++(0,-1) rectangle ++(1,1)};


\draw[-latex] (40,-4.3) --node[right] {$\mathsf{Top}_3$} ++(0,-1.5);


\fill[red] foreach \x in {1} {(34+\x,-7) rectangle ++(1,1)};

\fill[green] foreach \x in {1} {(34+\x,-6) -| ++(1,-1) --cycle};

\fill[yellow] foreach \x in {9} {(34+\x,-7) rectangle ++(1,1)};

\fill[red] foreach \x in {9} {(34+\x,-7) |- ++(1,1) --cycle};

\fill[green] foreach \x in {9} {(34+\x,-6) -| ++(1,-1) --cycle};

\fill[green] foreach \x in {11} {(34+\x,-7) rectangle ++(1,1)};

\draw (34,-7) ++(0,1) foreach \x in {0,1,2,...,11} {++(0,-1) rectangle ++(1,1)};

				\draw[semithick,-latex] (12.25,-2.5) -- ++(1.875,0) -- ++(0,1.1) -- ++(1.875,0);
				\draw[thick,-latex] (29.45,-6.7) -- ++(1.875,0) -- ++(0,5.3) -- ++(1.875,0);
			\end{scope}
		\end{tikzpicture}
  \caption{The constant-length sparsification method is applied across three adjacent satellites, with Satellite 1 being the farthest from the sink.}
  \label{fig:cons_spar}
	\end{figure*}

\subsection{Sparse Incremental Aggregation } \label{sec: Sparse_IA}

In this approach, termed \textit{\cgls{sia}}, the $k$th satellite updates its gradient $\vec{g}_k^n$ by combining it with the sparsification error from the previous iteration, $\vec{e}_k^{n-1}$, as follows: $\tilde{\vec{g}}_k^n \leftarrow D_k \vec{g}_k^n + \vec{e}_k^{n-1}$. This step, known as error-feedback~\cite{Sourav2024sparse}, plays a crucial role in integrating the residuals from earlier iterations, which helps accelerate convergence. Then, the $\text{Top}_Q$ is applied to $\tilde{\vec{g}}_k^n$, resulting in $\bar{\vec{g}}_k^n \leftarrow \text{Top}_Q(\tilde{\vec{g}}_k^n)$. Further, the residual for next iteration is updated as $\vec{e}_k^n \leftarrow \tilde{\vec{g}}_k^n - \bar{\vec{g}}_k^n$. 

In the aggregation phase, satellite $k$ aggregates the incoming gradient $\vec{\gamma}_{k-1}^t$ with its sparsified gradient $\bar{\vec{g}}_k^n$, updating it as $\vec{\gamma}_k^t \leftarrow  \vec{\gamma}_{k-1}^t + \bar{\vec{g}}_k^n$, before transmitting it to the next satellite $k+1$. The overall process is outlined in \cref{alg:spars_inc_agg}~\cite{Sourav2024sparse}. Further, the aggregation is performed in the common indices and otherwise values are forwarded with indices.

\begin{algorithm}[t]
	\caption{Sparse incremental aggregation at satellite $k$} \label{alg:const}
	\begin{algorithmic}[1]
		\StartIA{$\vec g_k^n$, $\vec\gamma_{k-1}^n$}
  \State Error feedback $ \tilde{\vec{g}}_k^n \leftarrow D_k \vec{g}_k^n + \vec e_k^{n-1}$  \label{alg:spars_inc_agg}
		\State Sparsification $\bar{\vec {g}}_k^n \gets \text{Top}_{{Q}}(\tilde{\vec g}_k^n)$  \label{alg:const:2}
  \State Update error $\vec e_k^n \gets \tilde{\vec{g}}_k^n - \bar{\vec {g}}_k^n$ \label{alg:const_len}
		\State Aggregation $\vec\gamma_k^n \gets \vec\gamma_{k-1}^n + \bar{\vec {g}}_k^n$ \label{alg:const:3}
		
		\EndIA{$\vec\gamma_k^n$}
	\end{algorithmic}
\end{algorithm}

To illustrate, consider \cref{fig: Sparse_IA}, and the process begins with satellite 1, which is farthest from the sink satellite. After applying the Top$_3$ operation on its error-compensated gradient, satellite 1 transmits its sparse gradient $\bar{\vec{g}}_1^n$ to satellite 2. For example, assume satellite 1 retains the values corresponding to indices $\{4, 8, 10\}$, where the indices are numbered starting from 1. Similarly, satellite 2 performs the Top$_3$ operation on its error-compensated gradient, resulting in values at indices $\{2, 4, 12\}$. Satellite 2 aggregates the values of the common indices (in this case, index 4). It then forwards this updated value of index 4 together with received values from satellite 1 in indices $\{8, 10\}$ and its own gradient values in indices $\{2, 12\}$ to the satellite 3. Satellite 3 repeats this process, combining and forwarding sparse gradients as the aggregate moves closer to the sink satellite.

In \cgls{sia}, the outgoing data budget for each satellite depends on the overlap between the support of the incoming sparse aggregate and its own sparse gradient. When the supports align, the outgoing budget remains unchanged from that of the previous satellite, allowing for the benefits of incremental aggregation. However, as noted in~\cite{Sourav2024sparse, Razmi2024On-Board}, after applying the Top$_Q$ operation with a low $Q$, the gradient supports become nearly uncorrelated. As a result, \cgls{sia} primarily involves forwarding both the sparse gradient from the previous satellite and the satellite's own sparse gradient, with aggregation occurring in only a few indices. This leads to a growing data budget as the aggregate moves toward the sink satellite, ultimately reducing the efficiency of incremental aggregation. Additionally, due to the variability in transmission budget requirements across clients, predetermining a fixed budget becomes challenging.


\subsection{Constant-Length Sparse Incremental Aggregation}
To leverage the benefits of incremental aggregation and stabilize the increasing budget in \cgls{sia}, an intuitive approach is to apply Top$_Q$ after combining each satellite’s gradient with the incoming values from the previous satellite. In this setup, the satellite aggregates only values at shared indices, while retaining its own gradient values at other indices. This ensures a fixed transmission budget of \( Q \) parameters per satellite, with a budget of \( (\omega + \lceil \log_2 n_d \rceil)Q \). Accordingly, this approach is termed \textit{\cgls{cl_sia}}.

Here, at each $n$th iteration, satellite $k$ begins by updating its gradient to account for errors from the previous iteration, resulting in an error-compensated gradient: $
\tilde{\vec{g}}_k^n \leftarrow D_k \vec{g}_k^n + \vec{e}_k^{n-1}.
$
Next, satellite $k$ aggregates this error-compensated gradient with the sparse aggregate $\vec{\gamma}_{k-1}^n$ received from the preceding satellite $k-1$, yielding
$
\tilde{\vec{\gamma}}_k^n \leftarrow  \vec{\gamma}_{k-1}^n + \tilde{\vec{g}}_k^n.
$
The Top$_Q$ is then applied to this aggregate, resulting in 
$
\vec{\gamma}_k^n \leftarrow \text{Top}_Q (\tilde{\vec{\gamma}}_k^n).
$
This sparse $\vec{\gamma}_k^n$ is then forwarded to the next satellite $k+1$, along with the associated indices. Finally, satellite $k$ updates its sparsification error 
$
\vec{e}_k^n \gets \tilde{\vec{\gamma}}_k^n - \vec{\gamma}_k^n,
$
as outlined in \cref{alg:const_len} \cite{Sourav2024sparse}.

To illustrate, consider the \cref{fig:cons_spar}, which begins with satellite 1, that is farthest from the sink satellite. Satellite 1 first applies the Top$_3$ operation on its error-compensated gradient, retaining only the most significant components. It then transmits this sparse gradient to satellite 2. For instance, suppose satellite 1 retains values corresponding to the indices $\{4, 8, 10\}$. Upon receiving this sparse gradient, satellite 2 aggregates with its own error-compensated gradient, combining values at common indices. It then applies the Top$_Q$ operation on this aggregate to produce a sparse result, with selected indices, say, $\{2, 4, 12\}$. Therefore, utilizing benefits of the incremental aggregation.
Similarly, satellite 3, upon receiving the aggregate from satellite 2 at indices $\{2, 4, 12\}$, performs the same steps to produce its own sparse aggregate, as the aggregate moves closer to the sink satellite.

\begin{algorithm}[t]
	\caption{Constant-length sparse incremental aggregation at satellite $k$} 
	\begin{algorithmic}[1]
		\StartIA{$\vec g_k^n$, $\vec\gamma_{k-1}^n$}
  \State Error feedback $ \tilde{\vec{g}}_k^n \leftarrow D_k \vec{g}_k^n + e_k^{n-1}$  \label{alg:const:21}
		\State Aggregation $ \bar{\vec{\gamma}}_k^n \leftarrow \tilde{\vec{g}}_k^n +\vec \gamma_k^{n}$  \label{alg:const:22}
		\State Sparsification $\vec\gamma_k^n \gets \text{Top}_{Q}(\tilde{\vec \gamma}_k^n)$ \label{alg:const:4}
		\State Update error $\vec e_k^n \gets \tilde{\vec\gamma}_k^n - \vec\gamma_k^n$ \label{alg:const_len}
		\EndIA{$\vec\gamma_k^n$}
	\end{algorithmic}
\end{algorithm}

Note that, the \cgls{cl_sia} leverages incremental aggregation by maintaining a consistent budget at each satellite. However, as aggregate values accumulate across satellites, combining multiple previous gradients, it's possible that the actual Top$_Q$ values of a satellite’s own gradient might not be retained in the final Top$_Q$ operation after aggregation. This is more likely to happen as the aggregate moves toward the sink satellite.
That leads to higher individual sparsification error at the satellites, which are closer to the sink satellite, potentially increasing the overall error and degrading convergence.

\section{Numerical Evaluation} \label{sec:numeval}

\begin{figure}[t]
	\centering
		\begin{tikzpicture}
              \begin{axis} [
yminorgrids = true, 
legend entries = {\cgls{isl} no-Spars, \cgls{isl} SIA, \cgls{isl} CL-SIA, no-\cgls{isl} SIA.},
xlabel={Time [h]},
ylabel={Test Accuracy},
				grid=major,
				minor x tick num = 4,
				minor y tick num = 1,
xmin = 0,
xmax = 32,
ymin = 0,
ymax = 0.95,
grid = major,
width=0.99*\axisdefaultwidth,
height=1*\axisdefaultheight,
legend cell align={left},
legend pos=south east,
legend style={font=\small}
				]
				\pgfplotstableread[col sep=comma]{figures/FedISL_FedNonISL_spars_acc_time_wo_included.csv}\tbl


                \addplot[black, thick] table [x=Time_FedISL_Sync_Bremen_wo_spars, y=Acc_FedISL_Sync_Bremen_wo_spars] {\tbl};
  
				\addplot[blue, thick] table [x=Time_FedISL_Sync_Bremen_SIA, y=Acc_FedISL_Sync_Bremen_SIA] {\tbl};
                   
				\addplot[green, thick] table [x=Time_FedISL_Sync_Bremen_CL, y=Acc_FedISL_Sync_Bremen_CL] {\tbl};
                    
				\addplot[brown, thick] table [x=Time_FedNonISL_Sync_Bremen_SIA, y=Acc_FedNonISL_Sync_Bremen_SIA] {\tbl};
    
			\end{axis}
   \end{tikzpicture}%
     \caption{Accuracy with respect to time with sparsification ratio $q=0.01$ for a constellation with $K=40$ satellites, where the \cgls{gs} is located in Bremen, Germany. Note that SIA and CL-SIA stand for sparse incremental aggregation and constant-length sparse incremental aggregation, respectively.}
     \label{fig:ACC-IA}
\end{figure}
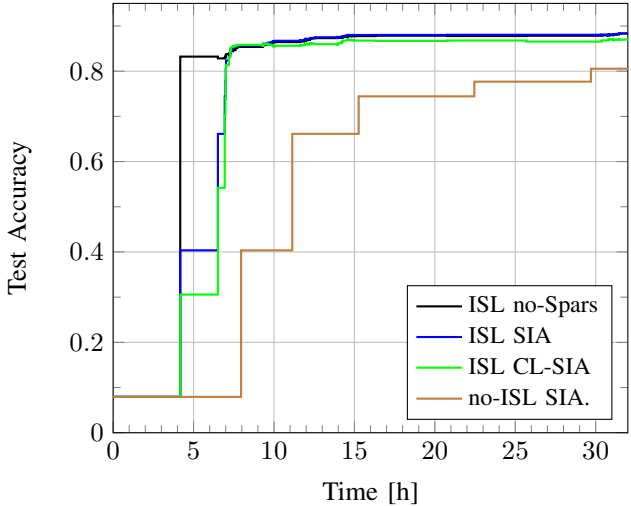

\begin{figure}[t]
		\begin{tikzpicture}
              \begin{axis} [
                        thick,
					font = {\small},
					xlabel={Number of satellites in an orbital plane},
					ylabel={Total transmitted data [\si{\mega\bit}/global iteration]},
					grid=major,
					minor x tick num = 1,
					minor y tick num = 1,
					scaled y ticks = base 10:-6,
					ytick scale label code/.code={},
					no markers,
					legend columns = 4,
					legend cell align=left,
					legend pos = {north west},
					legend image post style = {very thick},
         		width=.99*\axisdefaultwidth,
            	height=1*\axisdefaultheight,
				]
				\pgfplotstableread[col sep=comma]{figures/Spars_Acc_MNIST.csv}\tbl

				\addlegendimage{legend image with text=SIA}
				\addlegendentry{}
				\addlegendimage{legend image with text=CL-SIA}
				\addlegendentry{}
    			\addlegendimage{legend image with text=no-SIA}
				\addlegendentry{}
				\addlegendimage{empty legend}
				\addlegendentry{}    

				\addplot[blue] table [x=round, y=SIA_001] {\tbl};
                    \addlegendentry{}
				\addplot[green] table [x=round, y=CL_001] {\tbl};
                    \addlegendentry{}
				\addplot[red] table [x=round, y=wo_IA_001] {\tbl};
                  \addlegendentry{}
                  \addlegendimage{empty legend}
				\addlegendentry{$q=0.01$}
    
			\addplot[blue, dashed] table [x=round, y=SIA_01] {\tbl};
                \addlegendentry{}
			\addplot[green, dashed] table [x=round, y=CL_01] {\tbl};
                \addlegendentry{}
			\addplot[red, dashed] table [x=round, y=wo_IA_01] {\tbl};
                \addlegendentry{}
                \addlegendimage{empty legend}
			\addlegendentry{$q=0.1$}
			\end{axis}
        \end{tikzpicture}
        \caption{Total transmitted data with respect to the number of the satellites in an orbital plane, with the sparsification ratios of $q=0.01$ and $q=0.1$. Note that SIA and CL-SIA stand for sparse incremental aggregation and constant-length sparse incremental aggregation, respectively.}
        \label{fig:absolute_data_size_mnist}
\end{figure}
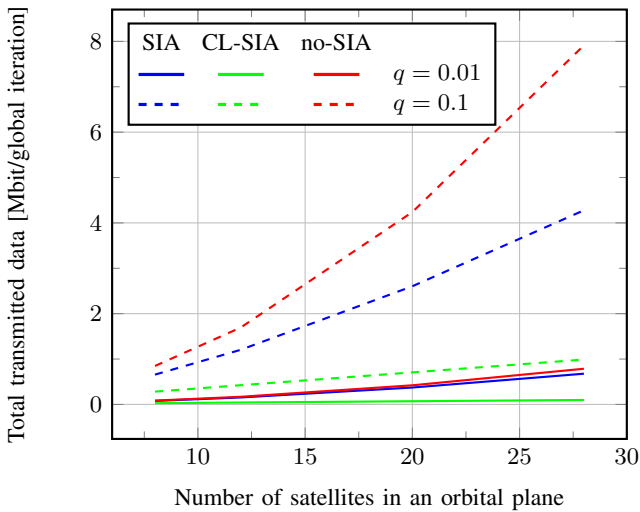

We present the performance of the proposed sparse incremental aggregation on a Walker star constellation with $K= 40$ satellites located in $P=5$ evenly spaced orbital planes at inclination $\ang{85}$ at an altitude $h_p=2000 \si{km}$. Each orbital plane consists of $8$ equidistant satellites. The \cgls{ps} located in Bremen, Germany.

For communication between satellites or between the \cgls{gs} and the satellites, we set the transmission power on $P_t = \SI{40}{\dBm}$, and both the transmitter and receiver antennas gains on \SI{32.13}{\dBi}. Communication occurs over a channel with a bandwidth of $B = \SI{500}{\mega\Hz}$. The carrier frequency is set to $f_c = \SI{20}{\giga\Hz}$, and the receiver noise temperature is $T = \SI{354}{\kelvin}$ \cite{leyva2022ngso}.

We evaluate the \cgls{ml} performance using a logistic regression model with $n_d = 7850$ trainable parameters on the MNIST dataset, which consists of grayscale images of handwritten digits ranging from 0 to 9 \cite{MNIST}. The data samples are distributed evenly and randomly across the satellites.

\cref{fig:ACC-IA} shows the test accuracy over wall clock time for \cgls{ia} with and without sparsification, as well as with and without \cglspl{isl}. In the case without \cglspl{isl} (no-ISL SIA), each satellite, in every iteration, sparsifies its gradients after training and transmits them to the \cgls{gs} during its visits. In this regard, the \cgls{gs} must wait to receive the sparsified gradient parameters individually from each satellite. The sparsification ratio is set on $q=0.01$. As it is seen, both SIA and CL-SIA, when combined with the \cgls{isl} algorithm, achieve higher accuracy in a shorter time compared to the scenario without \cglspl{isl}. This improvement occurs because, in the absence of intra-orbit \cglspl{isl}, the \cgls{gs} must wait to receive or transmit parameters from all satellites. Moreover, as observed, the case without sparsification achieves higher accuracy initially for several hours. However, after that period, the performance with (ISL SIA) and without (ISL no-Spars) sparsification converges. A noteworthy point is that \cgls{sia} shows slightly better performance than CL-SIA, which is attributed to the transmission of more data.

\Cref{fig:absolute_data_size_mnist} illustrates the total transmission data required to collect the gradients within a single orbital plane for both the SIA and CL-SIA algorithms. We evaluate the scenario with varying the number of satellites from 8 to \num{28} in the orbit, using sparsification ratios of $q=0.01$ and $q=0.1$. With the CL-SIA algorithm, we observe a substantial reduction in data load as the amount of transmitted data per hop remains fixed at its minimum possible value. In contrast, with the SIA algorithm, as the probability of having non-zero and non-overlapping entries increases after each hop, the communication load becomes significantly higher. Incorporating \cgls{ia} in each hop leads to a notable reduction in the communication load.

\section{Conclusions} \label{sec:Conclusions}
We considered satellites equipped with intra-orbit \cglspl{isl}. Using these \cglspl{isl}, satellites collectively forward model parameters within each orbit, either transmitting them to or receiving them from the \cgls{gs}. The relaying of model parameters to the \cgls{gs} relies on in-network aggregation. The system's performance is further improved by implementing advanced sparsification algorithms in aggregation step to optimize bandwidth usage. By utilizing the algorithms proposed in \cite{Sourav2024sparse}, we could significantly reduce the communication load for satellite constellations.

\bibliography{IEEEtrancfg,IEEEabrv,references}
\end{document}